\documentclass[10pt]{article}
\usepackage{graphicx}
\usepackage{pstcol,times,graphicx,graphics,pstricks,pst-node,pst-coil}
\begin{document}
\date{}

\title{Dynamical Casimir effect with Robin boundary conditions}

\author{B. Mintz\footnote{mintz@if.ufrj.br} , C. Farina\footnote{farina@if.ufrj.br} ,
P.A. Maia Neto\footnote{pamn@if.ufrj.br} $\,$ and R. Rodrigues\footnote{robson@if.ufrj.br}\\
\\
Instituto de F\'\i sica - Universidade Federal do Rio de Janeiro\\
Caixa Postal 68528 - CEP 21941-972, Rio de Janeiro, Brasil.\\
\\}
\bigskip
\maketitle
\begin{abstract}
{We consider a massless scalar field in 1+1 dimensions that satisfies a Robin boundary condition
at a non-relativistic moving boundary. Using the perturbative approach introduced by
Ford and Vilenkin, we compute the total force on the moving boundary. In contrast to what happens for the Dirichlet and Neumann boundary conditions, in addition to a dissipative part, the force acquires also a dispersive one.
Further, we also show that with an appropriate choice for the mechanical frequency of the moving boundary it is possible to turn off the vacuum dissipation almost completely.}
\end{abstract}




The interaction between  a physical system and a material plate (or cavity in general) in its surroundings has a
long history.  In 1948, Casimir and Polder \cite{CasimirPolder48} computed for the first time the retarded interaction energy between a neutral but polarizable atom and a perfectly conducting wall. At this same year, Casimir \cite{Casimir48} predicted the attraction between two neutral parallel conducting plates due to the shift caused by the plates in the energy of the radiation field in vacuum state. Casimir's result may be considered the first problem worked out in detail of the so called cavity QED. Since then, a lot of work has been done on the Casimir effect, see for instance the reviews
\cite{Plunien86,LivroMilonni,Mostepanenko97,BordagMohideenMostepanenko,LivroMilton,ArtigoMilton} 
and references therein (for other phenomena of cavity QED, see \cite{Haroche,Berman}).

However, the interaction between a quantum field and a material plate is quite complicated. Hence, as a first approximation, it is common to simulate this interaction by imposing an idealized boundary condition on the field . The most familiar conditions are Dirichlet and Neumann ones. A less familiar, but not less important condition is the so called Robin boundary condition, defined for a scalar field by
\begin{equation}\label{RobinBC}
\phi\vert_{\partial{\cal R}}=\beta\frac{\partial\phi}{\partial n}\vert_{\partial{\cal R}}\, ,
\end{equation}
where $\partial{\cal R}$ is the boundary of the system under study, $\frac{\partial\phi}{\partial n}$ means 
$\hat{\bf n}\cdot\nabla\phi$, with $\hat{\bf n}$ being a unitary vector normal to the boundary
and $\beta$ is a parameter with dimension of length that can assume any value in the interval $[0,\infty)$. Robin BC have the nice property of interpolating continuously Dirichlet and Neumann ones. From (\ref{RobinBC}), we immediatly see that for $\beta=0$ we have Dirichlet BC and for $\beta\rightarrow\infty$ we have Neumann BC.

In this work, we discuss some consequences of using Robin BC in the context of the Dynamical Casimir effect. However, before starting our calculations, we shall make a few comments about this kind of BC.
Robin BC already appear in a natural way in classical physics. For instance, when we solve problems in classical electromagnetism in the presence of spherical conducting shells the radial functions satisfy Robin BC with particular values of parameter $\beta$. Another nice example, still in the context of classical physics, is the problem of a vibrating string
subjected to a tension $T$ with two massless rings at its ends which may slide without friction along vertical rods and are coupled to springs of constants $\kappa_1$and $\kappa_2$, respectively, as indicated in figure \ref{CordacomMolas}:


\begin{figure}[!h]
\begin{center}
\newpsobject{showgrid}{psgrid}{subgriddiv=1,griddots=10,gridlabels=6pt}
\begin{pspicture}(-5,-1.7)(8,2.3)

\psset{arrowsize=0.2 2} \psset{unit=0.7}

\psline[linewidth=.05,linestyle=dashed]{->}(-5,0)(10,0)
\rput(-3.3,0.5){\Large${\cal O}$}
\rput(7.4,0.5){\Large$a$}
\rput(9.5,-0.7){\Large${\cal X}$}
\psline[linewidth=.1](-4,1.0)(-4,3)
\psline[linewidth=.1](-4,-3)(-4,0.9)
\psline[linewidth=.1](8,1.0)(8,3)
\psline[linewidth=.1](8,-3)(8,0.9)
%
\psellipse[linewidth=0.6mm](-4,1)(0.22,0.10)
\psellipse[linewidth=0.6mm](8,1)(0.22,0.10)
\pscoil[coilarm=.22,coilwidth=.45,coilheight=1.2](-4,-3.2)(-4,0.9)
\pscircle[linewidth=1mm](-4,-3){0.15}
\psline[linewidth=.03]{<->}(-3,-3)(-3,0)
\rput(-2.5,-1.5){\large$\ell_0$}
\rput(-4.7,-1.5){\Large$\kappa_1$}
\pscoil[coilarm=.22,coilwidth=.45,coilheight=1.2](8,-3.2)(8,0.9)
\pscircle[linewidth=1mm](8,-3){0.15}
\rput(7.2,-1.5){\Large$\kappa_2$}
%
\pscurve[linewidth=0.75mm](-3.8,1)(-3.5,1.3)(-3.0,1.6)(-2.5,1.7)(-2.0,1.7)(-1.5,1.5)(-1.0,1.1)
(-0.5,0.5)(-0.2,0.0)(0.0,-0.35)(0.3,-0.75)(0.8,-1.1)(1.0,-1.2)(1.3,-1.3)
(1.6,-1.35)(1.7,-1.35)(2.0,-1.35)(2.3,-1.3)(2.7,-1.2)(3.0,-1.1)(3.5,-0.8)(4.0,-0.35)
(4.26,0)(4.5,0.4)(4.7,0.75)(4.9,1.0)(5.0,1.15)(5.3,1.4)(5.8,1.65)(6.3,1.73)(6.8,1.7)(7.3,1.5)(7.8,1.0)
\psline[linewidth=.4mm]{->}(-3.8,1.05)(-2.0,2.8)
\psline[linewidth=.4mm]{<-}(6.1,2.9)(7.8,1.05)
\rput(-3.4,2.0){\large$T$}
\rput(7.4,2.0){\large$T$}
\psline[linewidth=.4mm]{->}(-4.5,1.0)(-4.5,-1.0)
\psline[linewidth=.4mm]{->}(8.5,1.0)(8.5,-1.0)

\end{pspicture}
\end{center}
\caption{Elastic supports at $x=0$ and $x=a$ give rise to Robin BC.}
\label{CordacomMolas}
\end{figure}
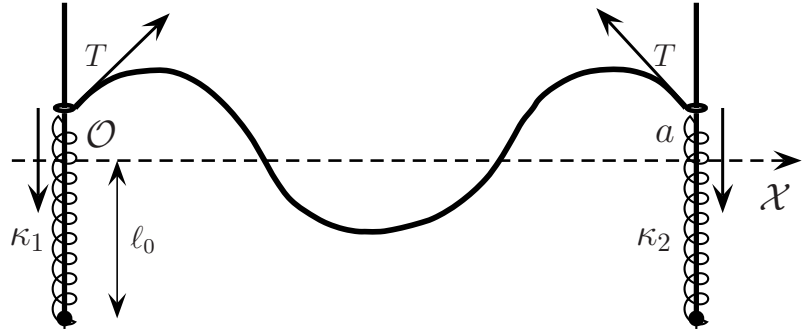
%
%

Assuming small inclinations $\bigl(\vert\frac{\partial y}{\partial x}\vert <<1\bigr)$, application of  Newton's Second Law to both massless rings  gives
\begin{equation} 
y\vert_{x=0} = \frac{T}{\kappa_1} \frac{\partial y}{\partial x}\vert_{x=0}
\;\;\;\;\;\mbox{and}\;\;\;\;\;\;
 y\vert_{x=a} = -\frac{T}{\kappa_2} \frac{\partial y}{\partial x}\vert_{x=a}
\end{equation}

The fact that Robin BC simulates an elastic support at the boundary has been pointed out in the literature \cite{GhenZhou}. Though the reflection at a fixed boundary where the wave satisfies a Robin BC is complete, there is some kind of time delay caused by a bulk/boundary dynamics. In other words, the reflection coefficient can be written as
$R=e^{i\phi(k)}$ (note that $\vert R\vert=1$), where $k$ is the wavenumber of the incident wave and hence there will be a phase shift between the incident and reflected waves. This gives a qualitative explanation for the surface terms that appear in connection with Robin BC in quantum field theory
\cite{AlbuquerqueCavalcanti,Fulling2003,KennedyCritchleyDowker,RomeoSaharian}. 
Total energy  (string plus  surface terms) is conserved, but there is a 
\lq\lq bulk/boundary{\rq\rq} exchange, so that the energy of the string itself is not conserved:
$$
\frac{d}{dt}\int_0^a\Biggl[\frac{1}{2}\mu\left(\frac{\partial y}{\partial t}\right)^2 +  \frac{1}{2}T\left(\frac{\partial y}{\partial x}\right)^2\Biggr] dx
= -\left\{
 \kappa_1 y(0,t)\frac{\partial y}{\partial t}(0,t) +
 \kappa_2 y(a,t)\frac{\partial y}{\partial t}(a,t)\right\}\; .
$$

Robin BC are also useful for phenomenological models that describe penetrable
surfaces \cite{MostepanenkoTrunov}. In fact, for some particular cases, these conditions 
can simulate the plasma model for real metals. It is not difficult to show that  
for frequencies much smaller than the plasma frequency, $\omega<<\omega_P$, a small value 
of $\beta$ plays the role of $1/\omega_P$ ($c=1$). In other words, under such assumptions, $\beta$ is proportional to the plasma wavelength, which is directly related to the penetration depth of the field.

Recently, Robin BC have been studied in many different contexts, namely: 
Bondurant and Fulling \cite{BondurantFulling} discussed in detail the Green's functions of the wave, heat and Schr\"odinger equations under Robin BC; Albuquerque and Cavalcanti \cite{AlbuquerqueCavalcanti} and Albuquerque \cite{Albuquerque2005} analized the one-loop renormalization of a $\lambda\phi^4$ theory under these conditions; Minces and Rivelles
used them in the context of AdS/CFT correspondence; Solodukihn \cite{Solodukhin} studied 
upper bounds for the ratio between entropy and energy of systems constrained by Robin BC; 
heat kernel coefficients were studied by Bordag {\it et al} \cite{BordagFalomirEtAl}, Fulling 
\cite{Fulling2003} and Dowker \cite{Dowker2005}; and a very detailed calculation of the
static Casimir effect with Robin BC was made by Romeo and Saharian \cite{RomeoSaharian}. 
It is worth mentioning that Robin BC may give rise to restauring Casimir forces between two parallel plates, once parameters $\beta$ at each plate are appropriately chosen.

However, since the pioneering paper by Moore \cite{Moore70} on radiation reaction forces on moving boundaries,
 Robin BC have never been considered explicitly in the context of the dynamical Casimir effect (as far as the authors
 know). It is our purpose here to make this kind of calculation in a simple model, namely, we shall consider a massless
scalar field $\phi$ in 1+1 dimensions subjected to a Robin BC at one non-relativistic moving
boundary. The main motivation is the following: it has been shown that for
Dirichlet  \cite{FullingDavies,FordVilenkin,JaekelReynaudQO92} and
Neumann BC \cite{AlvesFarinaPAMN} the linear susceptibilities are equal and purely imaginary,
\begin{equation}\label{ChiDN}
\chi^D(\omega) = \chi^N(\omega) = i\frac{\hbar \omega ^{3}}{6\pi c^2}\, .
\end{equation}
These susceptibilities lead to purely dissipative forces on the moving boundary:
\begin{equation}\label{ForcasDN}
\delta F^D(t)=\delta F^N(t)=\frac{\hbar}{6\pi c^2}\frac{d^3}{dt^3}\delta {q}(t)\, ,
\end{equation}
where $\delta q(t)$ is the position of the moving boundary at instant $t$.

For more general BC see Jaekel and Reynauld \cite{JaekelReynaudQO92,JaekelReynaud92} and for 3+1 calculations see references  \cite{PAMN93-94}. Since Robin BC interpolates continuously Dirichlet and Neumann ones we are led to make the following
questions: what happens to the force for the interpolating BC? Will it still be a purely
dissipative one? In what follows we shall answer these questions.

Besides the assumption of a non-relativistic motion for the boundary, we shall also 
suppose that the boundary has a prescribed motion with a small amplitude,
$\delta q(t)$ being its position at time $t$. Hence, we assume that
\begin{equation}\label{Assumptions}
|\delta \dot{q}(t)|\ll c\;\;\;\; \mbox{and}\;\;\;\; |\delta q(t)|\ll c/\omega_0\; ,
\end{equation}
where $\omega_0$ corresponds to the typical mechanical frequency. Therefore, we need to solve
the wave equation for que quantum field, $\partial^2\phi (t,x)=0$, with $\phi$ satisfying
a Robin BC at the moving boundary, which, in the co-moving frame, is written as
\begin{equation}\label{BCCo-movingFrame}
\frac{\partial\phi^{\,\prime}}
{\partial x^{\,\prime}}(t^{\,\prime},x^{\,\prime})\vert_{Bound.} = \frac{1}{\beta}\phi^{\,\prime}(t^{\,\prime},x^{\,\prime})\vert_{Bound.} \; .
\end{equation}
The corresponding BC in the laboratory frame is given by
\begin{equation}\label{BCLabFrame}
\left[\frac{\partial}{\partial x} + \delta\dot q(t)\frac{\partial}{\partial t}\right] 
\phi(t,x)\vert_{x=\delta q(t)} = 
\frac{1}{\beta}\phi(t,x)\vert_{x=\delta q(t)}\; ,
\end{equation}
where we  neglected terms of ${\cal O}(\delta\dot q^2/c^2)$). Using the perturbative approach 
of Ford and Vilenkin \cite{FordVilenkin} we write
\begin{equation}\label{FV}
\phi (x,t)=\phi _{0}(x,t)+\delta \phi (x,t)\, ,  
\end{equation}
where $\phi_0$ is the solution with a static boundary at $x=0$, which is given by
\begin{equation}
\phi_0(t, x)= \int_0^\infty 
\frac{d\omega}
{\sqrt{\omega(1+\omega^2\beta^2)\pi}}
\Bigl[\sin(\omega x) + \omega\beta\cos(\omega x)\Bigr]
\Bigl[ a(\omega) e^{-i\omega t} +
 a^\dagger(\omega) e^{i\omega t} \Bigr]\, .
\end{equation}
and $\delta\phi$ corresponds to the contribution generated by the movement of the boundary. This perturbation satisfies the wave equation $\partial^2 \delta \phi (x,t)=0$ with the following BC:
\begin{equation}
\frac{\partial\delta\phi}{\partial x}(t,0)-
\frac{1}{\beta}\delta\phi(t,0)=
\delta q(t)\left[\frac{1}{\beta}
\frac{\partial\phi_0}{\partial x}(t,0)-
\frac{\partial^2\phi_0}{\partial x^2}(t,0)\right] - 
\delta \dot q(t)
\frac{\partial\phi_0}{\partial t}(t,0)\, ,
\end{equation}
where we discarded terms of ${\cal O}(\delta q^2)$. The total force on the boundary is given by:
\begin{equation}
\delta F(t)=\langle 0|T^{11}\bigl(t,\delta q^{+}(t)\bigr) - 
T^{11}\bigl(t,\delta q^{-}(t)\bigr)|0\rangle \; ,
\label{c_forca_def}
\end{equation}
where $T^{11}(t,x)=-\frac{1}{2}\biggl\{(\partial_{x}\phi )^{2}(t,x)+
(\partial_{t}\phi)^{2}(t,x)\biggr\}\, . $
Substituting  $\phi=\phi_o+\delta\phi$:
\begin{eqnarray}
\delta F(t)\!\!\!\!&=&\!\!\!\! -\frac{1}{2}\langle 0\vert
\Bigg(\Bigl\{(\partial _{x}\phi_{0})(t,\delta q^+(t)),
(\partial _{x}\delta \phi )(t,\delta q^+(t))\Bigr\} + 
 \nonumber\\
&+& 
\!\!\!\Bigl\{(\partial _{t}\phi_{0})(t,\delta q^+(t)),
(\partial _{t}\delta \phi )(t,\delta q^+(t))\Bigr\} - 
 \Bigl[\delta q^+(t)\rightarrow\delta q^-(t)\Bigr]\Biggr)\vert 0\rangle\; +\; {\cal O}(\delta\phi^2)\; ,
\nonumber
\end{eqnarray}
In the last equation $\bigl\{..,..\bigr\}$ means  anticomutator and terms involving only the non-perturbed field $\phi_0$ disappear. Now, we expand around $x=0$ and keep only first order terms. One may also show that the total force is twice the force on each side. With these facts in mind, we get
\begin{eqnarray}
&\delta& \!\!\!\!\!\! F(t)= -\frac{1}{2}\langle 0\vert\Biggl(
\Bigl\{(\partial _{x}\phi_{0})(t,0^+),
(\partial _{x}\delta \phi )(t,0^+)\Bigr\} + 
\nonumber\\
&{\;}&\;\;\;\;\;\;\; +\;\;\;
\Bigl\{(\partial _{t}\phi_{0})(t,0^+),
(\partial _{t}\delta \phi )(t,0^+)\Bigr\} -  
\Bigl[0^+\rightarrow 0^-\Bigr]\Biggr)\vert 0\rangle\nonumber\\
&=&
\!\!\!-\langle 0\vert
\Bigl\{(\partial _{x}\phi_{0})(t,0^+),
(\partial _{x}\delta \phi )(t,0^+)\Bigr\} \!+\! 
\Bigl\{(\partial _{t}\phi_{0})(t,0^+),
(\partial _{t}\delta \phi )(t,0^+)\Bigr\}\vert 0\rangle\; ,\nonumber
\end{eqnarray}
Denoting by $\delta{\cal F}(\omega)$, $\delta\Phi(\omega,x)$ and $\delta Q(\omega)$ the
time Fourier transforms of $\delta F(t)$, $\delta\phi(t,x)$ and $\delta q(t)$, respectively,
it is straightforward to show that  
\begin{eqnarray}
\delta{\cal F}(\omega)&=&-\int\frac{d\omega^{\,\prime}}{2\pi}\Biggl(
\langle 0\vert\Bigl\{\partial_x\Phi_0(\omega-\omega^{\,\prime},0),
\partial_x\delta\Phi(\omega^{\,\prime},0)\Bigr\} -\nonumber\\
 &-&
(\omega-\omega^{\,\prime})\omega^{\,\prime}
\Bigl\{\Phi_0(\omega-\omega^{\,\prime},0),\delta\Phi(\omega^{\,\prime},0)\Bigr\}
\Biggr)\vert 0\rangle\; .\nonumber
\end{eqnarray}
Hence, we must solve the equation $(\partial _{x}^{2}+\omega ^{2})
\delta \Phi (x,\omega)=0$ with the BC (this is condition (\ref{BCLabFrame}) translated to 
the Fourier space):
\begin{eqnarray}
\partial_x\delta\Phi(\omega^{\,\prime}, 0)
-\frac{1}{\beta}\delta\Phi(\omega^{\,\prime},0)\; &=&\; \frac{1}{\beta}
\int\frac{d\omega^{\,\prime\prime}}{2\pi}\partial_x\Phi_0(\omega^{\,\prime\prime},0)\delta Q(\omega^{\,\prime}-\omega^{\,\prime\prime})+\nonumber\\\nonumber\\
&+&
\int\frac{d\omega^{\,\prime\prime}}{2\pi}
\omega^{\,\prime}\omega^{\,\prime\prime}
\Phi_0(\omega^{\,\prime\prime},0)
\delta Q(\omega^{\,\prime}-\omega^{\,\prime\prime})\, .\nonumber
\end{eqnarray}
However, $\delta\Phi(\omega,x)$ satisfies a second order differential
equation, which means that we shall need an extra condition. A natural choice is to consider only the solutions for $\delta\phi(t,x)$ which describe perturbations getting away from the boundary:
\begin{equation}
\delta \Phi (\omega,x)\!=\!sgn(x)\frac{1}{i\,\omega}
\partial_x\delta \Phi (\omega,0)e^{i\omega |x|}
\;\;\;\;
\Longrightarrow
\;\;\;\;
\delta \Phi (\omega,0^\pm)\!=\pm \frac{1}{i\,\omega}
\partial_x\delta \Phi (\omega,0)\, .\nonumber
\end{equation}
Last equations allow us to express $\delta\Phi(\omega,0)$ and 
$\partial_x\delta\Phi(\omega,0)$ in terms of the static field. 
The resulting expressions, when substituted in $\delta{\cal F}(\omega)$, give
\begin{eqnarray}
\!\!\!\!&{\;}&\!\!\!\delta {\cal F}(\omega) = \int\frac{d\omega^{\,\prime}}{2\pi}\left(\frac{\beta\omega^{\,\prime}}
{i+\beta\omega^{\,\prime}}\right)
\int\frac{d\omega^{\,\prime\prime}}{2\pi}\delta 
Q(\omega^{\,\prime}-\omega^{\,\prime\prime})\times
\nonumber\\
\times
\Biggl(\!\!\!&-&\!\!\!\frac{1}{\beta}\langle 0\vert
\Bigl\{\partial_x\Phi_0(\omega -\omega^{\,\prime},0),
\partial_x\Phi_0(\omega^{\,\prime\prime},0)\Bigr\}\vert 0\rangle - \nonumber\\
&+&
\, \mbox{$...$}\,\; + \,
(\omega-\omega^{\,\prime})(\omega^{\,\prime}-\omega^{\,\prime\prime})\omega^{\,\prime\prime} 
\langle 0\vert 
\Bigl\{\Phi_0(\omega-\omega^{\,\prime},0), 
\Phi_0(\omega^{\,\prime\prime},0)\Bigr\}\vert 0\rangle
\Biggr)\; ,\nonumber
\end{eqnarray}
where $...$ means other  (though analogous) correlators. However, all correlators
in the previous expression are connected by the field equation and Robin BC to the following one:
$C_{0}(\omega _{1,}\omega _{2})=\langle 0|\bigl\{\Phi _{0}(\omega_{1},0),
\Phi _{0}(\omega _{2},0)\bigr\} |0\rangle$, which involves only the non-perturbed field.
A straightforward calculation leads to
\begin{equation}\label{Correlator}
C_0(\omega_1,\omega_2) =
\frac{4\pi\beta^2}
{(1+\omega_1^2\beta^2)}\vert\omega_1\vert\delta(\omega_1 + \omega_2)\, .\nonumber
\end{equation}
With the aid of this correlator, we write  $\delta{\cal F}(\omega)$ in the form
\begin{equation}
\delta {\cal F}(\omega )=:\chi (\omega )\;\delta Q(\omega )\;,
\label{c_chi}
\end{equation}
where the real and imaginary parts of the susceptibility $\chi(\omega)$ are identified as
\begin{equation}\label{ReChi}
{\cal R}e \chi(\omega) = -\frac{\omega\beta}{\pi}
\int_{-\infty}^\infty\! d\omega^{\,\prime}\;\frac{\omega^{\,\prime}
\vert\omega^{\,\prime}\!\!-\omega\vert\bigl[ 1 +\beta^2\omega^{\,\prime}
(\omega^{\,\prime}-\omega)\bigr]}
{\bigl[\beta^2(\omega^{\,\prime}-\omega)^2+1\bigr](\beta^2{\omega^{\,\prime}}^2 + 1)}
\end{equation}
and
\begin{equation}\label{ImChi}
{\cal I}m \chi(\omega)\! = \frac{1}{\pi}
\!\!\int_{-\infty}^{\infty}\! 
d\omega^{\,\prime} \frac{\omega^{\,\prime}\vert\omega^{\,\prime}\!\! -\omega\vert
\left\{ 1 + 2\beta^2\omega^{\,\prime}(\omega^{\,\prime} - \omega) + 
\beta^4{\omega^{\,\prime}}^2(\omega^{\,\prime} - \omega)^2\right\}
}
{\bigl[\beta^2(\omega^{\,\prime} - \omega)^2 + 1\bigr] 
(\beta^2{\omega^{\,\prime}}^2 + 1)}\, .
\end{equation}
Before we proceed, it is interesting to check some limits. Taking the limits $\beta=0$ (Dirichlet BC) or $\beta\rightarrow\infty$ (Neumann BC) in the above expressions,  we obtain
\begin{equation}
{\cal R}e \chi(\omega) \;\longrightarrow\; 0 
\;\;\; \mbox{and}\;\;\;
{\cal I}m \chi(\omega)\;\longrightarrow\;
\frac{i }{\pi }\int_{-\infty }^{+\infty }
{d\omega^{\,\prime }}\,\omega ^{\,\prime }|\omega -\omega ^{\,\prime }|\, .\nonumber
\end{equation}
As anticipated,  the susceptibility is purely imaginary in these limits. In order to
perform the integration for ${\cal I}m\chi(\omega)$ we need a regularization prescrition.
In this case, a very natural way of doing that is to write the integral in the form
\begin{equation}\label{c_regularizacao}
\left\{\int_{-\infty }^{+\infty }d\omega ^{\,\prime }{\omega ^{\,\prime }}|\omega
-\omega ^{\,\prime }|\right\}^{reg}
=
\lim\limits_{L\rightarrow \infty }\left(
\int_{-L}^{0}+\int_{0}^{\omega }+\int_{\omega }^{\omega +L}\right) d\omega
^{\,\prime }\omega ^{\,\prime }|\omega -\omega ^{\,\prime }|\, .
\end{equation}
Note that the first and third integral on the right hand side of the above equation cancel out. 
Therefore, we are left with
\begin{equation}
\chi (\omega )=i\frac{\hbar }{\pi }\int_{0}^{\omega }{d\omega ^{\,\prime }}%
\,\omega ^{\,\prime }|\omega -\omega ^{\,\prime }| =
i\frac{\hbar \omega ^{3}}{6\pi }\; ,
\end{equation}
which leads to the well known forces already written  in (\ref{ForcasDN}).
For later convenience, it is worth emphasizing that the total work is given by
\begin{equation}\label{TrabalhoTotal}
\int_{-\infty}^{+\infty} F(t)\delta \dot q(t)\, dt
=-\frac{1}{\pi}\int_0^\infty d\omega\,\omega\,
{\cal I}m\,\chi(\omega)\vert\delta Q(\omega)\vert^2\; .\nonumber
\end{equation}
Note that, for Dirichlet and Neumann BC, ${\cal I}m\chi(\omega)>0$ for
$\omega>0$, so that the force is always dissipative. 

Using a regularization prescription analogous to that described above
in equations (\ref{ReChi}) and (\ref{ImChi}),  the contributions for
the integrals coming from the intervals $(-\infty,0)$ and $(\omega,\infty)$
will cancel each other and we are left with integrals from $0$ to $\omega$.
Performing the remaining integrais, we obtain:
$$
{\cal R}e \chi(\omega)
= \frac{\omega}{\beta^2\pi}
\frac{-\beta\omega\log(\beta^2\omega^2+1) +\beta^3\omega^3 + 4\beta\omega -
2\,\mbox{tg}^{-1}(\beta\omega)(\beta^2\omega^2+2)}
{\beta^2\omega^2 + 4}
$$
$$
{\cal I}m \chi(\omega)
= \frac{\omega}{6\beta^2\pi}
\frac{\beta^4\omega^4 + 4\beta^2\omega^2 - 6\log(\beta^2\omega^2+1)(\beta^2\omega^2+2)
+12\beta\omega\,\mbox{tg}^{-1}(\beta\omega)}
{\beta^2\omega^2 + 4}\, .
$$
Expanding  the previous expressions appropriately, the first corrections to the Dirichlet and
Neumann cases can be obtained:

For $\beta\omega<<1$, we have
$$
{\cal R}e\chi(\omega) = -\frac{\omega^4}{6\pi}\beta + \frac{2\omega^6}{15\pi}\beta^3 + 
{\cal O}(\beta^5)\;\;\;\mbox{and}\;\;\;
{\cal I}m\chi(\omega) = \frac{\omega^3}{6\pi} - 
\frac{\omega^5}{6\pi}\beta^2 + {\cal O}(\beta^4)
$$

For $\beta\omega>>1$, we have
$$
{\cal R}e\chi(\omega) = -\frac{\omega^2}{\pi}\frac{1}{\beta} - 
\frac{\omega}{\beta^2} + {\cal O}(\beta^{-3})
\;\;\;\mbox{and}\;\;\;
{\cal I}m\chi(\omega) = \frac{\omega^3}{6\pi} - \frac{2\omega}{\pi}
\frac{\log(\beta\omega)}{\beta^2} + {\cal O}(\beta^{-4})
$$
Hence, the total force is given by
\begin{eqnarray}
\delta F(t) &=& \frac{1}{6\pi}\left\{\frac{d^3}{dt^3}\delta q(t) - 
\beta\frac{d^4}{dt^4}\delta q(t)\right\} +{\cal O}(\beta^2)\;\;\;\; 
(\beta\rightarrow 0) \nonumber\\
\delta F(t) &=& \frac{1}{\pi}\left\{\frac{1}{6}\frac{d^3}{dt^3}\delta q(t) - 
\frac{1}{\beta}\frac{d^2}{dt^2}\delta q(t)\right\} +{\cal O}(\beta^{-3})\;\;\;\; 
(\beta\rightarrow\infty)\nonumber
\end{eqnarray}

\begin{figure}[!hbt]
\begin{center}
\includegraphics[width=13cm]{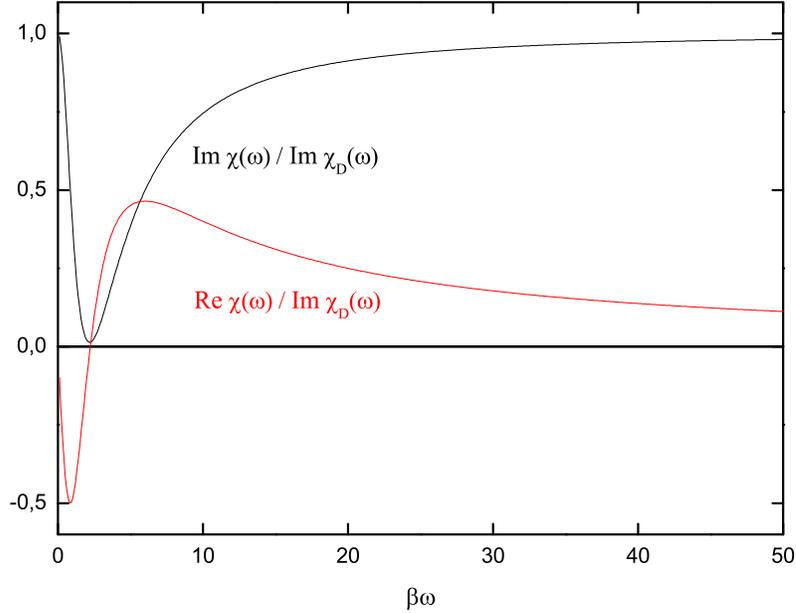}
\vskip -0.8 cm
\caption{Real and imaginary parts of $\chi(\omega)$, conveniently normalized by ${\cal I}m\chi_D(\omega)$, as
functions of $\beta\omega$.}
\end{center}
\label{DispersaoRobin}
\end{figure}

The behaviours of ${\cal R}e\chi(\omega)$ and ${\cal I}m\chi(\omega)$
are shown in figure \ref{DispersaoRobin}.
%


\vskip -1.0 cm


In this work, we computed the total force on a non-relativistic moving boundary in 1+1 dimensions
due to the vacuum fluctuations of a massless scalar field subjected to Robin BC. Dirichlet and
Neumann BC correspond to particular limits of our results (particular values of $\beta$).
 It is worth emphasizing that when Robin BC are used the susceptibility acquires a real part. The pronounced
 valley in the graph of ${\cal I}m\chi(\omega)/{\cal I}m\chi_D(\omega)$ leads to a quite interesting
 result: if $\delta Q(\omega)$ is peaked  around $\omega_0$, equation (\ref{TrabalhoTotal}) shows
 that, for any fixed $\beta$, there will be an appropriate choice of $\omega_0$ such that the dissipative effects
 on the boundary will be almost completely eliminated.  A natural sequence of this work is to compute
 the particle creation rate under the same  circumstances  as those assumed in this work.
 This problem is under study and the results will be published elsewhere.

 The authors thank CNPq and FAPERJ for financial support and S.A. Fulling for bringing
 reference \cite{GhenZhou} to our knowledge.


\footnotesize
\begin{thebibliography}{99}
%
\bibitem{CasimirPolder48} Casimir H B G and Polder D 1948 {\it Phys. Rev.} {\bf 73} 360
%
\bibitem{Casimir48} Casimir H B G 1948 {\it Proc. K. Ned. Acad. Wet.} {\bf 51} 793
%
\bibitem{Plunien86} Plunien G, M\"uller B and Greiner W 1986 {\it Phys. Rep.} {\bf 134} 88
%
\bibitem{LivroMilonni} Milonni P W in {\it The Quantum Vacuum: an introduction to quantum electrodynamics}, Academic Press, Inc., London (1994)
%
\bibitem{Mostepanenko97} Mostepanenko V M and Trunov N N 1997 {\it The Casimir Effect and its Applications}, Osford Science Publications, Oxford
%
\bibitem{BordagMohideenMostepanenko} Bordag M, Mohideen U and Mostepanenko V M  2001, 
{\it Phys. Rep.}{\bf 353} 1 
%
\bibitem{LivroMilton} Milton K A 2001 {\it The Casimir Effect: Physical Manifestations of Zero-Point Energy}, World Scientific, Singapore
%
\bibitem{ArtigoMilton} Milton K, J. Phys. A 2004 
%
\bibitem{Haroche} Serge Haroche, in  Fundamental Systems in Quantum Optics, Les Houches Summer School (1991), J. Dalibard, J.M. Raymond and J. Zinn-Justin, eds., Elsevier Science
%
\bibitem{Berman} {\it Cavity Quantum Eletrodynamics}, edited by Paul R. Berman, Academic Press, Inc., London 1994
%
\bibitem{GhenZhou} Chen G and Zhou J 1993 {\it Vibration and Damping in Distributed Systems}
vol {\bf I} (Boca Raton, FL:CRC) p15
%
\bibitem{AlbuquerqueCavalcanti} de Albuquerque L C and Cavalcanti R M 2004 {\it J. Phys. A} {\bf 37} 7039
%
\bibitem{Fulling2003} Fulling S A 2003 {\it J. Phys. A} {\bf 36} 6857
%
\bibitem{KennedyCritchleyDowker} Kennedy G, Critchley R and Dowker J S 1980 {\it Ann. Phys. (NY)} {\bf 125} 346
%
\bibitem{RomeoSaharian} Romeo A and Saharian A A 2002 {\it J. Phys. A} {\bf 35} 1297
%
\bibitem{BondurantFulling} Bondurant J D and  Fulling S A 2005 {\it J. Phys. A: Math. Gen.} {\bf 38} 1505
%
\bibitem{AsoreyRosaLeo} Asorey M, da Rosa F S and  Carvalho L B 2004, presented in the {\bf XXV ENFPC}, Caxamb\'u, Brasil
%
\bibitem{DeutschCandelas} Deutsch D and Candelas P 1979 {\it Phys. Rev. D}{\bf 20} 3063
%
\bibitem{MostepanenkoTrunov} Mostepanenko V M and Trunov N N 1985 {\it Sov. J. Nucl. Phys.} {\bf 45} 818
%
\bibitem{Solodukhin} Solodukhin S N 2001 {\it Phys. Rev. D}{\bf 63} 044002
%
\bibitem{MincesRivelles} Minces P and Rivelles V O 2000 {\it Nucl. Phys. B} {\bf 572} 651
%
\bibitem{BordagFalomirEtAl} Bordag M, Falomir H, Santangelo E M and Vassilevich D V 2002 {\it Phys. Rev. D} {\bf 65} 064032
%


\bibitem{Dowker2005} Dowker J S 2005 {\it math.SP/0409442 v4}
%
\bibitem{Albuquerque2005} de Albuquerque L C 2005 {\it hep-th/0507019 v1}
%

%
\bibitem{Moore70} Moore G T 1970 {\it Math. Phys.} {\bf 11} 2679
%
\bibitem{FullingDavies} Fulling S A and Davies P C W 1976 {\it Proc. R. Soc. London A} {\bf 348} 393
%
\bibitem{FordVilenkin} Ford L H and Vilenkin A 1982 {\it Phys. Rev. D} {\bf 25} 2569
%
\bibitem{JaekelReynaudQO92} Jaekel M T and Reynaud S 1992 {\it Quant. Opt.} {\bf 4} 39 .
%
\bibitem{AlvesFarinaPAMN} Alves D T, Farina C and Maia Neto P A 2003 {\it J. Phys. A} {\bf 36} 1133
%
\bibitem{JaekelReynaud92} Jaekel M T and Reynaud S 1992 {\it J. Phys. I (France)} {\bf 2} 149
%
\bibitem{PAMN93-94} Maia Neto P A and Reynaud S 1993 {\it Phys. Rev. A} {\bf 47} 1639;
Maia Neto P A 1994 {\it J. Phys. A} {\bf 27} 2167
%


\end{thebibliography}
\end{document}